\RequirePackage{fix-cm}
\documentclass[smallextended]{svjour3}       
\smartqed  
\usepackage{graphicx}
\usepackage{hyperref}
\usepackage{breakcites}
\usepackage{amsmath,nicefrac,booktabs}
\usepackage{amsfonts}
\usepackage{amssymb}
\usepackage{makeidx}
\usepackage{graphicx,float,subcaption}
\usepackage[margin=3cm]{geometry}
\usepackage{parskip}
\usepackage{pdflscape}
\usepackage{rotating}
\begin{document}

\title{Building a Dynamical Network Model from Neural Spiking Data
}
\subtitle{Application of Poisson Likelihood}


\author{Ozgur DORUK         \and
        Kechen Zhang 
}


\institute{R.O. DORUK \at
              Atilim University \\
              Tel.: +90-312-5868000\\
              \email{resat.doruk@atilim.edu.tr}           
           \and
           K. Zhang \at
              Johns Hopkins School of Medicine
}

\date{Received: date / Accepted: date}

\maketitle

\begin{abstract}
Research showed that, the information transmitted in biological neurons is encoded in the instants of successive action potentials or their firing rate. In addition to that, in-vivo operation of the neuron makes measurement difficult and thus continuous data collection is restricted. Due to those reasons, classical mean square estimation techniques that are frequently used in neural network training is very difficult to apply. In such situations, point processes and related likelihood methods may be beneficial. In this study, we will present how one can apply certain methods to use the stimulus-response data obtained from a neural process in the mathematical modeling of a neuron. The study is theoretical in nature and it will be supported by simulations. In addition it will be compared to a similar study performed on the same network model.
\end{abstract}

\section{Introduction}
\label{sec:intro}
Modeling of neurons go back to the mid of 1900s. The well known Hodgkin-Huxley (HH) model is derived to explain the quantitative behavior of the electrochemical activities in the squid giant axon \cite{hodgkin1952quantitative}. After this revolutionary development numerous studies followed such as Morris- Lecar \cite{morris1981voltage}, Fitzhugh - Nagumo \cite{fitzhugh1961impulses} and Hindmarsh - Rose \cite{hindmarsh1982model,hindmarsh1984model} models. The models developed after HH either adds extra details (such as calcium channel dyanmics) or simplify the overall model. Models targeting the simplification do this either by explaining a major function and discarding the other features or lumping all channels into a single variable. For example \cite{morris1981voltage} defines the activation mechanism with one single recovery variable (though there are lots of biophysical parameters and variables still existing). On the other hand in models such as \cite{fitzhugh1961impulses,hindmarsh1982model,hindmarsh1984model} explains the behavioral details such as repetitive firing, bursting etc. Thus, they will not involve physical parameters or only have a very few of them. Depending on the research/application the existence of physical parameters in a neuron may or may not be necessary. In fact, one major criterion in building/selection of the neuron model is the answer to the question: How will one collect the data? In the development of models such as HH, the data is collected through a voltage-clamp \cite{pehlivan2009biyofizik} experiment. This is a standard method in electrophysiology. However, it requires the neuron to be isolated (or in vitro experiment). In an in-vivo application, placing an electrode to the neuron's membrane is risky as it will alter the operation of the cell (propagation of the action potential might be delayed etc.). That will be disastrous. Measurement without touching to the neuron in consideration is possible but one will not be able to measure the level of action potential. However, one can collect the individual peaks of successive action potentials thanks to the local current flows from the membrane. This will yield an array of time values which defines the locations of the peaks of the successive action potentials. This array is often called as neural spike train as each action potential is considered like a sudden voltage jump. We will not have continuous data collection here. We will only have timing information but no potential levels. At a first look one may think that this data has no meaning but the reality seems different. Studies such as \cite{shadlen1994noise} showed that this temporal data carries the actual coding rather than the action potential itself. In addition, it is also understood that the spike trains are not deterministic and they appear at different locations even the same stimulus is applied. It is discovered that the stochastic distributions of the spike timings obey an Inhomogeneous Poisson Process with the event rate being the neural firing rate of the neurons.

Being aware of that one might propose a generic model that can be trained using point process likelihood methods \cite{myung2003tutorial}. These are asymptotically efficient methods where the mean square error approaches the Cramer-Rao lower bound as the number of data samples increases. Knowing the fact that neural spiking obeys Poisson processes, one can implement a likelihood function from the Poisson's probability mass function. One can make use of the spike count for the evaluation. 

A similar study was performed by \cite{dimattina2011active} and \cite{dimattina2013adaptive}. Here, static feedforward neural network is fitted from the discrete neural spiking data. The parameter estimation procedure is based on the maximum a-posteriori (MAP) estimation technique \cite{murphy2012machine} which is known as an extension to the maximum likelihood method (ML). The model targets a stimulus-response relationship and includes a very few physical parameters. Though it is relatively easier to process a static neural network model, it lacks the features such as time dependentness which may not be adequate to express the behavior of a realistic neuron or neural network. In fact there is a dynamical version of this network which is called as continuous time recurrent neural network (CTRNN) \cite{beer1995dynamics}. This includes some nonlinear features existent in realistic neural networks and also has the advantage of universal approximation capability. In this work, we will work on a CTRNN type model. 

We can summarize what to be done in this research as follows:
\begin{enumerate}
	\item We will present an approach on how we can fit a model from a discrete neural spiking data.
	\item The study targets the testing of efficiency of the algorithms that is to be used in estimation of the parameters of the neuron.   
	\item The neural spike data will be generated from the firing rate through Poisson process simulation.
	\item The parameters of the model will be estimated using Maximum Likelihood Estimation and the probability mass function of the Inhomogeneous Poisson Process is used as a likelihood function. 
	\item The optimization of the likelihood function is performed through MATLAB's \texttt{fmincon} script. 
	\item The comparison of the findings with a study \cite{doruk2017fitting} using a different likelihood function will also be briefly presented. 
\end{enumerate}                        
\section{Theoretical Methods}
\subsection{Model of a Neuron}
\label{sec:model-neuron}
As in several physical processes neurons exhibit a highly nonlinear behavior. So one can express the stimulus response relationship of a neuron or a neural network as a generic nonlinear system:
\begin{align}\label{eq:modgen}
\begin{split}
\dot{x} &= f(x,I,p) \\
y &= h(x)
\end{split}
\end{align}
In the above equation $ I $ is the stimulus and $ y $ is the response. Here the state variable $ x $ represent the time dependentness of the neurons. This may be a membrane potential, firing rate or a dimensionless variable. \eqref{eq:modgen} can represent a single neuron or an average response of a group of neurons. For a successful representation, it is recommended that at least two state variables should exist in the model. In Section \ref{sec:intro} we stated that, we will model the neuron by a continuous time recurrent neural network. So a CTRNN in general can be shown as:
\begin{equation}\label{eq:ctrnn}
\tau_{i}\frac{dx_{i}}{dt}=-x_{i}+\sum_{j=1}^{n} W_{ij}g_{j}\left(x_{j}\right) +\sum_{k=1}^{m} C_{ik}I_{k}
\end{equation}  
Here $ x_i $ and $ I_k $ are the state and stimulus variables like in \eqref{eq:modgen}. $ W_{ij} $ is a weight parameter representing the synaptic connections between neurons $ i $ and $ j $. The term $ C_{ik} $ is a weight parameter between the stimulus and the neurons. The function ${g_j\left(x_j\right)}$ is a soft saturable function relating the firing rate response of the neuron to its dynamical variable $ x_j $. Mathematically it is:
\begin{equation}
g_{j}\left(x_{j}\right)=\frac{\Gamma_j}{1+\exp\left( -a_{j}\left(x_{j}-h_j\right)\right) }\label{eq:sigmoid-general}
\end{equation} 
The above is also called as a sigmoid function. Here $\Gamma_j$ is the maximum firing rate that the neuron $ j $ can produce. $ h_j $ and $ a_j $ are a soft threshold and a slope parameter for the same neuron respectively. $ \tau_i $ is a time constant parameter. This is the major physical parameter here. In this work, we will work on a two neuron CTRNN model. This model will have one excitatory and one inhibitory neuron. More truely speaking the collective behavior of a group of excitatory and inhibitory neurons is lumped into two neurons. Mathematically this will be:
\begin{align}\label{eq:2nr-model}
\begin{split}
\tau_e\dot{x}_e &= -x_e+w_{ee}g_e(x_e)-w_{ei}g_i(x_i)+c_eI \\
\tau_i\dot{x}_i &= -x_i+w_{ie}g_e(x_e)-w_{ii}g_i(x_i)+c_iI
\end{split}
\end{align} 
In the above, $ e $ and $ i $ stand for excitatory and inhibitory units , $ \tau_e $ and $ \tau_i $ are the time constants of the excitatory and inhibitory units respectively, $ w_{ee} $ is the self excitation weight for the excitatory neuron, $ w_{ii} $ is the self inhibition constant for the inhibitory neuron, $ w_{ei} $ is the synaptic coefficient that represent the synapse that inhibits the excitatory neuron, $ w_{ie} $ is the synaptic coefficient that represents the synapse that excites the inhibitory neuron. All those parameters are positive in value and their excitatory characteristics are determined by their signs in \eqref{eq:2nr-model} (positive for excitatory and negative for inhibitory). $ c_e $ ve $ c_i $ are the coefficients of interaction between stimulus and the neurons. $ g_e $ ve $ g_i $ functions are obtained by replacing $ j $ in \eqref{eq:sigmoid-general} by $ e $ and $ i $. In order to rewrite \eqref{eq:2nr-model} in the form of \eqref{eq:modgen} we need to transfer the time constants as reciprocal forms to right hand side as:
\begin{align}\label{eq:2nr-model-betali}
\begin{split}
\dot{x}_e &= -\beta_ex_e+\beta_e\left( w_{ee}g_e(x_e)-w_{ei}g_i(x_i)+c_eI\right)  \\
\dot{x}_i &= -\beta_ix_i+\beta_i\left( w_{ie}g_e(x_e)-w_{ii}g_i(x_i)+c_iI\right) 
\end{split}
\end{align}
where $ \beta_e=\frac{1}{\tau_e} $ and $ \beta_i=\frac{1}{\tau_i} $. The stimulus entering the model is $ I $ here. The response will be the firing rate of the excitatory neuron which is denoted by $ r_e $. Of course this is not a measurable variable but it reveals itself in the spiking response of the excitatory neuron. The relationship to the excitatory dynamic variable $ x_e $:
\begin{equation}\label{eq:re-xe}
r_e=g_e(x_e)
\end{equation} 

In the above equations $ x_e $ and $ x_i $ are dynamical variables representing the excitatory and inhibitory neurons. They do not have to represent any physical quantity or process. 
          
\subsection{Neural Spiking, Poisson Random Processes and Likelihood}
\label{sec:neur-spik-pois-proc}
\begin{figure}[H]
	\centering
	\includegraphics[scale=1]{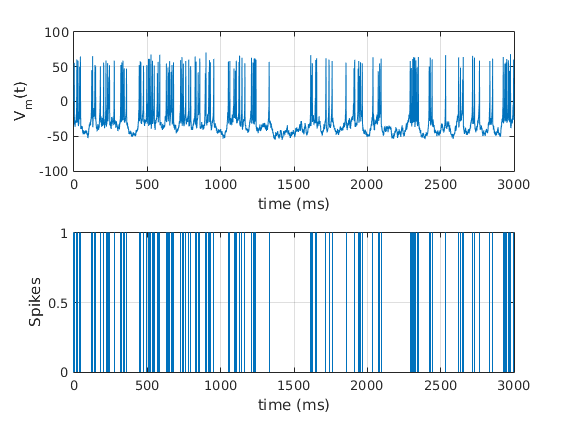}
	\caption{Action potentials and neural spiking: Peak points of individual action potentials taken as ones and the remaining regions correspond to a zero. This requires precision sampling as the spiking event can occur in a time bin of 1 ms or less.} \label{fig:act-pol-neur-spk}
\end{figure}

Neural spiking is the phenomena which occurs due to successive action potentials occurring in a neural transmission \cite{rieke1999spikes}. One can see an example in Figure \ref{fig:act-pol-neur-spk}. In this figure, the instant where the action potential is fired is recorded as '1' and '0' is recorded when the neuron is at rest state. This will be like a serial data recorded as a binary number (like RS-232 serial transmission).  
The transmitted information may be coded in the firing rate of the spikes \cite{kandel2000principles,adrian1926impulses}, their count \cite{forrest2014intracellular,forrest2014sodium} or timing \cite{dayan2003theoretical,butts2007temporal,singh2017consensus}. The mechanism may differ from neuron to neuron. We also reminded in Section \ref{sec:intro} that the spike trains such as the one shown in Figure \ref{fig:act-pol-neur-spk} is not deterministic and found to obey an Inhomogeneous Poisson Process \cite{shadlen1994noise}. 

In order to talk about the statistics of an Inhomogeneous Poisson Process one needs to write its probability mass function:
\begin{equation}
\mbox{Prob}\left[N\left(t+\Delta t\right)-N\left(t\right)=k\right]=\frac{e^{-\lambda} \lambda^{k}}{k!}
\label{eq:inhomogeneous-poisson}
\end{equation}
The above expression provides the probability of $ k $ number of events to occur in the interval $ {\left[t,t+\Delta t\right)} $. This probability depends on a critical parameter called as event rate $ \lambda $. In homogeneous Poisson process, this parameter is constant. In Inhomogeneous versions it will be time varying and equivalent to the neural firing rate $ r_e(t) $. Thus we will need to define the above in terms of the mean firing rate:
\begin{equation}
\lambda=\int_{t}^{t+\Delta t}r_e\left(\tau\right)d\tau
\end{equation}
In the neural spiking phenomenon, the number of spikes will be equivalent to the event count parameter which is $ k $. So one can determine the expected number of spikes by simulating the process defined by \eqref{eq:inhomogeneous-poisson}. As firing rate $ r_e $ depends on the state variable $ x_e $ one can also write the following:
\begin{equation}
\lambda(\theta)=\int_{t}^{t+\Delta t}r_e\left(\theta,\tau\right)d\tau
\end{equation} 
In the above we redefine firing rate as a function of the model parameters. That is:
\begin{equation}
\theta=\left[\theta_1, \ldots,\theta_8\right]
=\left[\beta_e,\beta_i,w_e,w_i,w_{ee},w_{ei},w_{ie},w_{ii}\right]\label{eq:theta-ctrnn-param}
\end{equation}
So we can say that, the probability of $ K_m $ spikes to occur in the interval $ {[0,T_m)} $ is given by:
\begin{equation}
p(K_m,\theta)=\frac{e^{-\lambda(\theta)} \lambda^{K_m}(\theta)}{K_m!}
\label{eq:inhomogeneous-poisson-thetali}
\end{equation} 
regarding the fact that firing rate generated by \eqref{eq:2nr-model-betali} and \eqref{eq:re-xe}. The above equation is also the likelihood function for trial $ m $. Statistically it is not enough to proceed with a single set of data and we often need multiple trials. In addition the stimulus $ I $ should be different for each trial (so $ I $ becomes $ I_m $). So, if we have $ M $ different stimuli and $ M $ corresponding different responses we can write the following:
\begin{equation}\label{eq:likelihood-pi}
p(K_1,K_2,\ldots,K_M,\theta)=\prod\limits_{m=1}^{M}\frac{e^{-\lambda_m(\theta)} \lambda_m^{K_m}(\theta)}{K_m!}
\end{equation}
In the above we have $ M $ different stimuli $ I_m $ and thus $ M $ different response $ r_e^m $ through the states $ x_e^m $. If $ r_e^m $ represents the firing rate of the trial $ m $:
\begin{equation}\label{eq:lambda-m}
\lambda_m(\theta)=\int_{0}^{T_m}r_e^m\left(\theta,\tau\right)d\tau
\end{equation}
is written. 
\eqref{eq:likelihood-pi} gives the joint likelihood of $ M $ independent neural spike data collection trials. In the optimization we generally prefer its logarithmic version as:
\begin{equation}\label{eq:likelihood-log}
L(K_1,K_2,\ldots,K_M,\theta)=\sum\limits_{m=1}^{M}\left( -\lambda_m(\theta)+{K_m}\ln[\lambda_m(\theta)]-\ln(K_m!)\right) 
\end{equation}   
The term $ K_m! $ may be neglected. One can write the estimate of parameter $ \theta $ in \eqref{eq:theta-ctrnn-param} as shown below:
\begin{equation}
\hat\theta_{ML}=\arg\max_{\theta} L(K_1,K_2,\ldots,K_M,\theta) 
\label{eq:mle-arg-max-log}
\end{equation}
The above optimization problem can be solved by MATLAB\textsuperscript{\textregistered} \texttt{fmincon} script. 
\subsection{Simulation of Poisson Processes and Spiking Data}    
\label{sec:sim-poiss}
In order to test the methodologies obtained in the last section one has to generate a spike train. The best way to achieve that is to obtain a set of timing events by simulating an Inhomogeneous Poisson Process (using \eqref{eq:inhomogeneous-poisson}) as a function of time dependent firing rate $ r_e $. This even better for the cases where the neuron model has a very few or no physical parameters. There are few different methods to simulate an Inhomogeneous Poisson Process by computational tools such as MATLAB. One feasible method when one has discrete time bins is the local Bernoulli approximation \cite{eden2008point}. Here one can assume that each spike is formed at a very narrow time bin such as 1 ms. Based on those one can say that:
\begin{enumerate}
	\item Assume that the firing rate of the neurons are given by $ r_e(t) $. 
	\item Suppose that $ \Delta t $ is an interval such that only one spike can appear in. 
	\item So, the probability of a spike to appear in the interval $ [t,t+\Delta t] $ will be $ r_e(t)\Delta t $.
	\item The probability of a spike not to exist in $ [t,t+\Delta t] $ interval will be $ 1-r_e(t)\Delta t $.  
\end{enumerate}  
From the above properties one can develop the following algorithm to generate a neural spike train:
\begin{enumerate}
	\item Divide the simulation interval $ [0,T_f] $ to $ \Delta t $ spaced discrete time bins. Now one will have $ N_f=\nicefrac{T_f}{\Delta t}+1 $ equally spaced time bins in the simulation interval. Note that $ \Delta t $ so small that only one spike can fit. 
	\item Suppose that current time instant is denoted by $ t_c $. In order to test whether there is any spike occurring at $ t_c $, first generate a uniformly distributed random variable $ x_{rand} $ in the range $ [0,1] $. In MATLAB this can be done by $ x_{rand} $=\texttt{unifrnd(0,1)}. 
	\item If $ r_e(t_c)\Delta t>x_{rand} $ fire a spike at $ t_c $.
	\item If $ r_e(t_c)\Delta t \leq x_{rand} $, nothing happens. 
	\item The steps up-to this point should be repeated for each time bin in $ [0,T_f] $. So one has $ N_f $ operations to process, but this should be fairly easy thanks to the vectorial computation capabilities of MATLAB.      
\end{enumerate}
\section{Example Application}
\label{sec:example}
\subsection{Definition of the Problem}
\label{sec:prob-definition}
In this section, we will present an example to demonstrate the theoretical information presented in Section \ref{sec:neur-spik-pois-proc}. We will attempt to estimate the parameters of \eqref{eq:2nr-model-betali} through the collected spike timing information (spike train). The parameters to be estimated are given in \eqref{eq:theta-ctrnn-param}. We can summarize the goals and procedures as follows:
\begin{enumerate}
	\item The model in \eqref{eq:2nr-model-betali} will be simulated using the nominal value of parameters in Table \ref{tab:nom-vals}. This will provide the true firing rate information that is to be encoded in the spike trains. 
	\item The method in Section \ref{sec:sim-poiss} will be used to simulate an Inhomogeneous Poisson Process to obtain the expected number and timings of the spikes. The event rate will be the firing rate obtained in the previous step. 
	\item The simulation will be repeated $ M $ times to obtain statistically adequate information. Each trial will involve a different stimulus. Information on the numerical value of $ M $ is given in Section \ref{sec:ex-scenario}. The generation of stimulus is described in Section \ref{sec:stimumuls-gent}. 
	\item After completion of data collection one can use joint likelihood function defined in \eqref{eq:likelihood-log}. The optimization problem is solved by MATLAB \texttt{fmincon} script. 
	\item The evaluation will be repeated a few times to see whether the algorithm works efficiently. Especially the convexity of the problem can only be understood this way. 
	\item The results will be presented as tables which reveals the mean value of estimated parameters, percent and mean square errors.  
\end{enumerate}
\begin{table}[H]
	\centering
	\caption{Nominal values of the parameters in \eqref{eq:2nr-model-betali}}\label{tab:nom-vals}
	\begin{tabular}{ccc}
		\toprule 
		Parameter & Unit & Nominal Value $\left(\theta\right)$ \\
		\midrule
		${\beta_e}$ & $ \nicefrac{1}{s} $ & ${50}$ \\
		\midrule
		${\beta_i}$ & $ \nicefrac{1}{s} $ & ${25}$ \\
		\midrule
		${w_e}$ & - & ${1.0}$ \\
		\midrule
		${w_i}$ & - & ${0.7}$ \\
		\midrule
		${w_{ee}}$ & - & ${1.2}$  \\
		\midrule
		${w_{ei}}$ & - & ${2.0}$ \\
		\midrule
		${w_{ie}}$ & - & ${0.7}$ \\
		\midrule 
		${w_{ii}}$ & - & ${0.4}$ \\
		\bottomrule 
	\end{tabular}
\end{table}
In the current problem it is assumed that $ g_e(x_e) $ and $ g_i(x_i) $ are known and their parameters are given in Table \ref{tab:g-param}. 
\begin{table}[H]
	\centering
	\caption{The values of parameters of the functions $ g_e(x_e) $ and $ g_i(x_i) $. They are assumed to be known and will not be estimated.}\label{tab:g-param}
	\begin{tabular}{cc}
		\toprule 
		Parametre & Değer \\ 
		\midrule 
		$ \Gamma_e $ & 100 \\ 
		\midrule
		$ a_e $ & 0.04 \\
		\midrule
		$ h_e $ & 70 \\
		\midrule 
		$ \Gamma_i $ & 50 \\ 
		\midrule
		$ a_i $ & 0.04 \\
		\midrule
		$ h_i $ & 35 \\ 
		\bottomrule
	\end{tabular}
\end{table}
\subsection{Formation of the Stimulus}
\label{sec:stimumuls-gent}
The stimulus is applied as an input $ I $ to the model. This may represent any physical exciter such as temperature, pressure, light etc. It can also be thought as the firing of the pre-synaptic neuron which enters as an stimulating input to the examined neuron or network. In view of mathematics, one has to define the input variable as a variable $ I(t) $. Concerning the fact that we are discussing sensory neurons we may think of a Fourier series describing the stimulus as:
\begin{equation}
I=\sum_{n=1}^{N}A_n \cos\left(\omega_{n}t+\phi_{n}\right)\label{eq:cosine-stimulus}
\end{equation}
In the above $ A_n $ is amplitude, $ \omega_n=2\pi f_0 n $ is the base frequency of each stimulus component and $ \phi_n $ is their phase. For multiple trials one can modify the above so that it is given for each trial:
\begin{equation}
I_m=\sum_{n=1}^{N_U}A_n \cos\left(\omega_{n}t+\phi_{n}^m\right)\label{eq:cosine-stimulus-multi}
\end{equation}
In the above $ m=1 \ldots M $. $ M $ is the number of trials as discussed in Section \ref{sec:prob-definition}.  
The stimulus $ I_m $ should be defined different in each case by assigning the phase randomly as a uniformly distributed number between $ (-\pi,\pi) $. This can be done in MATLAB by \texttt{phi=unifrnd(-pi,pi)}. 

\begin{figure}[H]
	\begin{subfigure}[b]{0.5\textwidth}
		\includegraphics[scale=0.5]{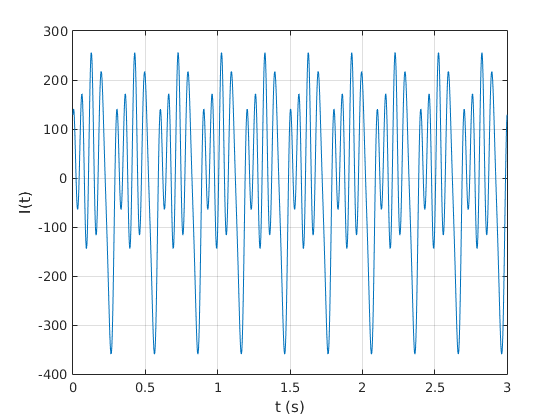}
		\caption{Stimulus $ I(t) $}
	\end{subfigure}
	\begin{subfigure}[b]{0.5\textwidth}
		\includegraphics[scale=0.5]{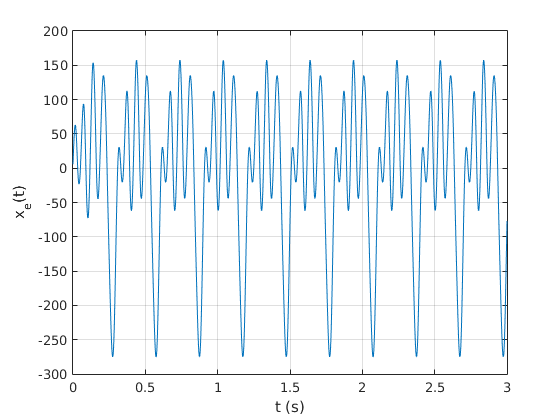}
		\caption{Excitatory Neuron Variable $ x_e(t) $}
	\end{subfigure}
	\begin{subfigure}[b]{0.5\textwidth}
		\includegraphics[scale=0.5]{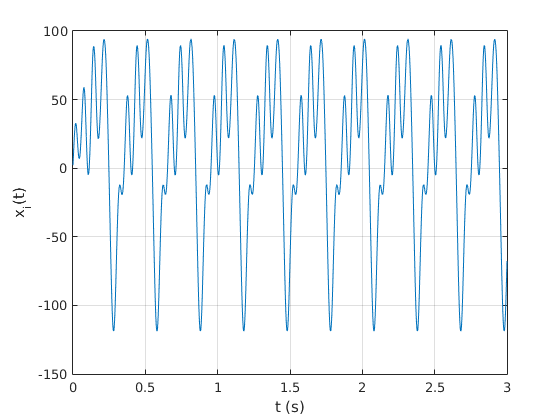}
		\caption{Inhibitory Neuron Variable $ x_i(t) $}
	\end{subfigure}
	\begin{subfigure}[b]{0.5\textwidth}
		\includegraphics[scale=0.5]{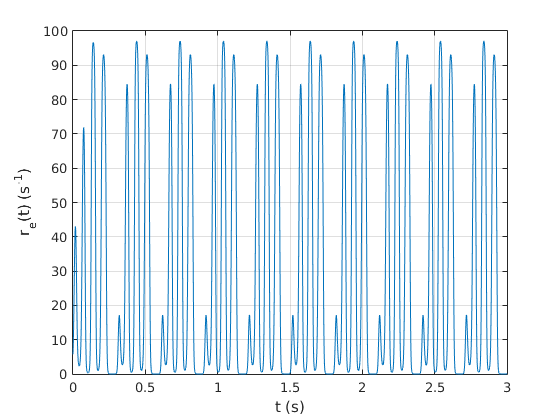}
		\caption{Firing Rate of the Excitatory Neuron $ r_e(t) $}
	\end{subfigure}
	\begin{subfigure}[b]{0.5\textwidth}
		\includegraphics[scale=0.5]{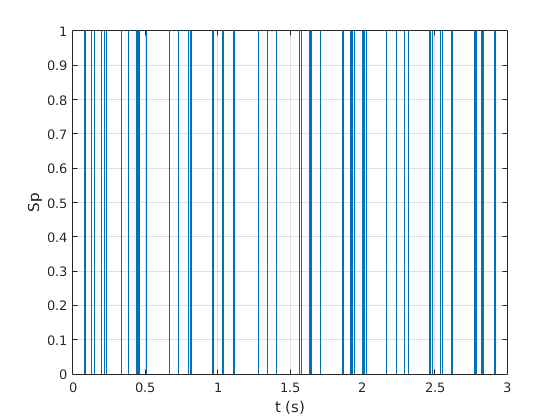}
		\caption{Excitatory Neuron Spike Train}
	\end{subfigure}
\caption{The response of the neural network in \eqref{eq:2nr-model-betali} with the nominal parameters given in Table \ref{tab:nom-vals} against stimulus defined in \eqref{eq:cosine-stimulus} configured with the $ N_U=5 $, $ f_0=3 $ Hz, $ A_n=100 $ and a random phase $ \phi_n $ parameters. }
\label{fig:our-neuron-demonst}
\end{figure}

Finally one will be able to see what happens when a stimulus is configured with $ N_U=5 $, $ f_0=3 $ Hz, $ A_n=100 $ and a random phase $ \phi_n $. That is presented in Figure \ref{fig:our-neuron-demonst}. 
\subsection{Example's Scenario} 
\label{sec:ex-scenario}   
In Table \ref{tab:sim-data} one can see the scenario associated with the example application. For comparison some of the parameters such as amplitude and sample size are also tried at different values. 
\begin{table}[H] 
	\centering
	\scriptsize
	\caption{The details associated with the scenario of the example application. Some parameters are multiple as they indicate the different cases examined for comparison.}
	\begin{tabular}{ccc}
		\toprule 
		Parameter & Notation & Value \\ 
		\midrule
		Duration of Simulation & $ T_f $ & 3 sec.\\ 
		\midrule
		Number of Repeats & $ M $ & 25,50,100,200,400 \\
		\midrule
		Size of Stimulus & $ N_U $ & 5,10,20,30,40,50 \\
		\midrule
		Optimization Algorithm & N/A & Interior-Point Gradient Descent (MATLAB) \\
		\midrule
		Number of Parameters & Size($ \theta $) & 8 \\
		\midrule
		Stimulus Amplitude & $ A_n $ & 25,50,100 \\
		\midrule 
		Base Frequency of Stimulus & $ f_0 $ & 3.333 Hz \\
		\midrule 
		Bin Size & $ \Delta t $ & 1 ms \\
		\bottomrule
	\end{tabular} 
	\label{tab:sim-data}
\end{table} 
\section{Presentation of the Results and Concluding Remarks}
\label{sec:results-presentation}
\subsection{General Evaluation}
\label{sec:gen-eval}  
\begin{figure}[H]
	\centering
	\includegraphics[scale=0.8]{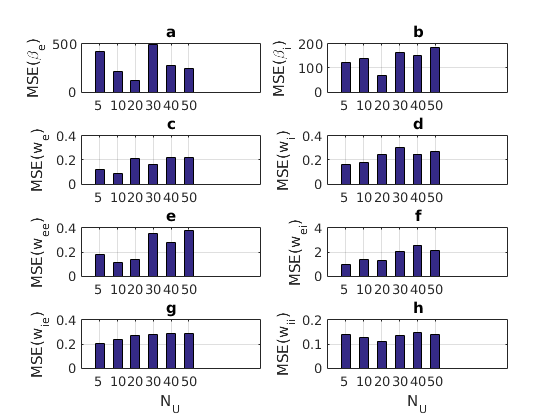}
	\caption{The variation of mean square errors (MSE) of each parameter against varying stimulus component size $ N_U $.  }
	\label{fig:mse-vs-nu}
\end{figure}

\begin{figure}[H]
	\centering
	\includegraphics[scale=0.8]{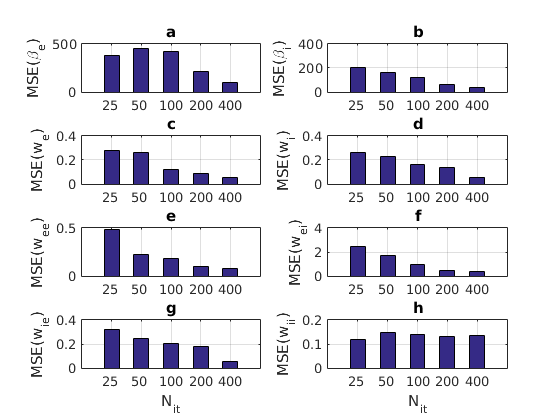}
	\caption{The variation of mean square errors (MSE) of each parameter against varying sample (or iteration) size $ M $.  }
	\label{fig:mse-vs-nit}
\end{figure}

In this section, one will be able to see the results obtained by the solution of the maximum likelihood estimation problem for the example presented in Section \ref{sec:example} using the likelihood function defined in \eqref{eq:likelihood-log}. The parameters to be estimated are given in \eqref{eq:theta-ctrnn-param}. The results are presented in two forms. The mean estimation results are presented in tabular form (\textbf{Table \ref{tab:results-finale}}) where as the variation of the mean square errors of estimation against stimulus component size $ N_U $ and the sample size $ M $ are presented separately as graphs (\textbf{Figure \ref{fig:mse-vs-nu} and \ref{fig:mse-vs-nit}}).

For most of the parameters (except $ w_{ii} $) the sample size $ M $ leads to decrease in mean square error. One can note that from \textbf{Figure \ref{fig:mse-vs-nit}} and also roughly from \textbf{Table \ref{tab:results-finale}}. The stimulus component size $ N_U $ does not seem to have a definable pattern. Here, it is recommended that large values of $ N_U $ are definitely unnecessary.  

\begin{sidewaystable}[htb!]

	\centering

	\caption{The results of application of the theory presented in Section \ref{sec:neur-spik-pois-proc} to the example problem in Section \ref{sec:example}. It is understood that the closeness of the mean values to the nominal parameters in Table \ref{tab:nom-vals} depends mainly on $ M $. Concerning the influence of the stimulus component size $ N_U $, one can not say that there is a definite pattern. This is aligned with the general properties of likelihood estimation.}\label{tab:results-finale}

	\footnotesize

	\begin{tabular}{cccccccccccc}

		\toprule

		Case & $ M $ & $ A_n $ & $ N_U $ & $ \hat{\beta}_e $ & $ \hat{\beta}_i $ & $ \hat{w}_e $ & $ \hat{w}_i $ & $ \hat{w}_{ee} $ & $ \hat{w}_{ei} $ & $ \hat{w}_{ie} $ & $ \hat{w}_{ii} $ \\

		\midrule

		1 & 25 & 25 & 5 & 55.533727 & 30.213046 & 0.921087 & 1.032323 & 1.732064 & 2.028027 & 0.848702 & 0.32529 \\

		2 & 25 & 50 & 5 & 56.608858 & 20.548860 & 1.029538 & 0.883940 & 1.349837 & 2.575930 & 0.804324 & 0.49538 \\ 

		3 & 25 & 100 & 5 & 57.188796 & 23.585112 & 1.070947 & 0.655189 & 1.517467 & 3.075277 & 0.662610 & 0.31510 \\

		4 & 50 & 25 & 5 & 50.313627 & 32.285748 & 0.973964 & 1.100521 & 1.699105 & 1.984025 & 0.781615 & 0.33855 \\

		5 & 50 & 100 & 5 & 56.597545 & 22.221102 & 1.097007 & 0.757587 & 1.361170 & 2.611427 & 0.772177 & 0.46856 \\

		6 & 100 & 25 & 5 & 55.119586 & 26.456978 & 0.806735 &  1.081219 & 1.609574 & 1.897204 & 0.694758 & 0.37113 \\

		7 & 100 & 100 & 5 & 57.238965 & 21.655775 & 0.913813 & 0.699147 & 1.370589 & 2.137143 & 0.877339 & 0.41435 \\

		8 & 200 & 100 & 5 & 51.998178 & 22.484430 & 0.951887 & 0.748606 & 1.260704 & 1.962527 & 0.790993 & 0.47072 \\

		9 & 400 & 100 & 5 & 50.205221 & 23.943961 & 0.964450 & 0.666517 & 1.298419 & 2.092121 & 0.797156 & 0.54510 \\

		10 & 100 & 100 & 10 & 53.484995 & 20.802459 & 1.008136 & 0.687495 & 1.268520 & 2.430492 & 0.645363 & 0.382041 \\
		11 & 100 & 100 & 20 & 48.692376 & 22.146853 & 1.153791 & 0.843028 & 1.206127 & 2.463636 & 0.586859 & 0.400202 \\
		12 & 100 & 100 & 30 & 60.642974 & 26.017975 & 0.918924 & 0.559221 & 1.602176 & 2.628880 & 0.772155 & 0.429427 \\
		13 & 100 & 100 & 40 & 53.158671 & 22.334121 & 1.168865 & 0.728257 & 1.416619 & 3.087837 & 0.758917 & 0.380532 \\
		14 & 100 & 100 & 50 & 55.625371 & 28.878888 & 1.100684 & 0.583886 & 1.556403 & 2.970991 & 0.859251 & 0.449933 \\
		\bottomrule

	\end{tabular}

\end{sidewaystable}

\subsection{Comparison with Another Likelihood Function}  
\label{sec:comparison} 
In this section we will compare the results obtained from this research with that of another similar research using a different likelihood function \cite{doruk2017fitting} which depends on the temporal locations of individual spikes (denoted as a set by $ S_m $) instead of just their count:
\begin{equation}
p\left(S_m,\theta\right)
=\exp\left(-\int_{0}^{T}r_e^{(m)}\!\left(t\right)dt\right)\prod_{k=1}^{K_m} r_e^{(m)}\!\left(t_{k}^{(m)}\right)
\label{eq:pSm}
\end{equation} 
In the compared study, the model and its parameters are same as that of \eqref{eq:2nr-model-betali} and the stimulus is also the same given by \eqref{eq:cosine-stimulus-multi}. Applying the same scenario as Table \ref{tab:sim-data} reveals that the work in \cite{doruk2017fitting} yields much smaller estimation mean square errors when considered the same sample size $ M $. In addition, smaller sample sizes in \cite{doruk2017fitting} generate much smaller estimation errors appear at smaller values of $ M $ than larger values of $ M $ in this study. The principal reason lying under that result should be the inclusion of the temporal locations of the individual spikes in the likelihood function \eqref{eq:pSm} from the collected spike trains. Of course this advantage required higher computational efforts and memory than the study in this research. Thus depending on the resources one can prefer the current work or \cite{doruk2017fitting}. However, the covariance of the estimation error seems much higher in the application of this paper.


\bibliographystyle{spbasic}      

\end{document}